\documentstyle[11pt,amssymb,epsf]{article}

\def\eq#1{(\ref{#1})}

\def\crampest{\medmuskip = 1mu plus 1mu minus 1mu}
\def\uncramp{\medmuskip = 4mu plus 2mu minus 4mu}
\def\ben{\begin{equation}}
\def\een{\end{equation}}

\let\a=\alpha    
    
  \let\n=\nu

\let\C=\Chi 
\let\la=\label  
  
\def\nn{\nonumber} \def\bd{\begin{document}} \def\ed{\end{document}}
\def\ds{\documentstyle} \let\fr=\frac \let\bl=\bigl \let\br=\bigr
\let\Br=\Bigr \let\Bl=\Bigl
\let\bm=\bibitem
\let\na=\nabla
\let\pa=\partial \let\ov=\overline
\newcommand{\be}{\begin{equation}}
\newcommand{\ee}{\end{equation}}
\def\ba{\begin{array}}
\def\ea{\end{array}}
\def\ft#1#2{{\textstyle{{\scriptstyle #1}\over {\scriptstyle #2}}}}
\def\fft#1#2{{#1 \over #2}}
\def\del{\partial}
\def\vp{\varphi}
\def\sst#1{{\scriptscriptstyle #1}}
\def\oneone{\rlap 1\mkern4mu{\rm l}}
\def\td{\tilde}
\def\wtd{\widetilde}
\def\ie{\rm i.e.\ }
\def\dalemb#1#2{{\vbox{\hrule height .#2pt
        \hbox{\vrule width.#2pt height#1pt \kern#1pt
                \vrule width.#2pt}
        \hrule height.#2pt}}}
\def\square{\mathord{\dalemb{6.8}{7}\hbox{\hskip1pt}}}
\newcommand{\ho}[1]{$\, ^{#1}$}
\newcommand{\hoch}[1]{$\, ^{#1}$}
\newcommand{\bea}{\begin{eqnarray}}
\newcommand{\eea}{\end{eqnarray}}
\newcommand{\ra}{\rightarrow}
\newcommand{\lra}{\longrightarrow}
\newcommand{\Lra}{\Leftrightarrow}
\newcommand{\ap}{\alpha^\prime}
\newcommand{\bp}{\tilde \beta^\prime}
\newcommand{\tr}{{\rm tr} }
\newcommand{\Tr}{{\rm Tr} }
\def\0{{\sst{(0)}}}
\def\1{{\sst{(1)}}}
\def\2{{\sst{(2)}}}
\def\3{{\sst{(3)}}}
\def\4{{\sst{(4)}}}
\def\5{{\sst{(5)}}}
\def\6{{\sst{(6)}}}
\def\7{{\sst{(7)}}}
\def\8{{\sst{(8)}}}
\def\n{{\sst{(n)}}}
\def\cA{{{\cal A}}}
\def\cF{{{\cal F}}}
\def\tV{\widetilde V}
\def\tW{\widetilde W}
\def\tH{\widetilde H}
\def\tE{\widetilde E}
\def\tF{\widetilde F}
\def\tA{\widetilde A}
\def\im{{{\rm i}}}
\def\tY{{{\wtd Y}}}
\def\ep{{\epsilon}}
\def\vep{{\varepsilon}}
\def\R{\rlap{\rm I}\mkern3mu{\rm R}}
\def\bD{{{\bar D}}}

\def\R{\rlap{\rm I}\mkern3mu{\rm R}}
\def\bD{{{\bar D}}}
\def\R{{{\Bbb R}}}
\def\C{{{\Bbb C}}}
\def\H{{{\Bbb H}}}
\def\CP{{{\Bbb C}{\Bbb P}}}
\def\RP{{{\Bbb R}{\Bbb P}}}
\def\Z{{{\Bbb Z}}}
\def\bA{{{\Bbb A}}}
\def\bB{{{\Bbb B}}}
\def\bC{{{\Bbb C}}}
\def\bR{{{\Bbb R}}}
\def\bD{{{\Bbb D}}}
\def\bE{{{\Bbb E}}}
\def\bZ{{{\Bbb Z}}}
\def\Re{{{\frak{Re}}}}
\def\Im{{{\frak{Im}}}}
\def\cosec{{\,\hbox{cosec}\,}}
\def\Gm{{\Gamma_{\!\! -}}}
\def\Gp{{\Gamma_{\!\! +}}}
\def\stan{{standard }}
\def\nonstan{{supernumerary }}

\def\cosech{{\hbox{cosech}}}

\def\etcyc{{\hbox{and cyclic}}}
\def\btheta{{\bar\theta}}

\newcommand{\tamphys}{\it Center for Theoretical Physics,
Texas A\&M University, College Station, TX 77843, USA}
\newcommand{\umich}{\it Michigan Center for Theoretical Physics,
University of Michigan\\ Ann Arbor, MI 48109, USA}
\newcommand{\upenn}{\it Department of Physics and Astronomy,\\
University of Pennsylvania, Philadelphia,  PA 19104, USA}
\newcommand{\SISSA}{\it  SISSA-ISAS and INFN, Sezione di Trieste\\
Via Beirut 2-4, I-34013, Trieste, Italy}

\newcommand{\mitchell}{\it George P. and Cynthia W.
Mitchell Institute for Fundamental Physics,\\
Texas A\&M University, College Station, TX 77843-4242, USA}

\newcommand{\newton}{\it Isaac Newton Institute for Mathematical Sciences,\\
20 Clarkson Road,  University of Cambridge,
Cambridge CB3 0EH, UK}

\newcommand{\ihp}{\it Institut Henri Poincar\'e\\
  11 rue Pierre et Marie Curie, F 75231 Paris Cedex 05}

\newcommand{\damtp}{\it DAMTP, Centre for Mathematical Sciences,
 Cambridge University\\  Wilberforce Road, Cambridge CB3 OWA, UK}
\newcommand{\itp}{\it Institute for Theoretical Physics, University of
California\\ Santa Barbara, CA 93106, USA}

\newcommand{\auth}{
H. L\"u,\hoch{*} C.N. Pope\hoch{*} and E. Sezgin\hoch{\dagger}}

\begin{document}
\begin{flushright}
\hfill{MIFP-02-04}\\
\hfill{
\bf hep-th/0212323}
\end{flushright}

\vspace{15pt}

\begin{center}

{\large {\bf $SU(2)$ Reduction of Six-dimensional $(1,0)$
Supergravity}}

\vspace{15pt}

\auth

\vspace{7pt}
{\hoch{\ddagger}\mitchell}

\vspace{75pt}

\underline{ABSTRACT}
\end{center}

\vspace{15pt}

    We obtain a gauged supergravity theory in three dimensions with
eight real supersymmetries by means of a Scherk-Schwarz reduction of
pure $N=(1,0)$ supergravity in six dimension on the $SU(2)$ group
manifold. The $SU(2)$ Yang-Mills fields in the model propagate, since they
have an ordinary kinetic term in addition to Chern-Simons
couplings. The other propagating degrees of freedom consist of a
dilaton, five scalars which parameterise the coset $SL(3,R)/SO(3)$,
three vector fields in the adjoint of $SU(2)$, and twelve spin $\ft12$
fermions. The model admits an AdS$_3$ vacuum solution. We also show
how a charged black hole solution can be obtained, by performing a
dimensional reduction of the rotating self-dual string of
six-dimensional $(1,0)$ supergravity.

{\vfill\leftline{}\vfill
\vskip 10pt \footnoterule
{\footnotesize \hoch{*}
Research supported in part by DOE grant DE-FG03-95ER40917
\vskip -12pt} \vskip 14pt
{\footnotesize \hoch{\dagger}
Research supported in part by NSF Grant PHY-0070964
\vskip -12pt}  \vskip  14pt
}

\pagebreak
\setcounter{page}{1}

\tableofcontents
\addtocontents{toc}{\protect\setcounter{tocdepth}{2}}
\newpage

\section{Introduction}

    The importance of supergravities in diverse dimensions in the
context of string theory has been widely appreciated for a long
time. However, the role of gauged supergravities has been appreciated
only relatively recent, and especially it has become more evident with
developments in the AdS/CFT correspondence. Supergravities are
referred to as ``gauged'' when their $R$-symmetry group, or any
subgroup thereof, is gauged. Those which have played role in the
AdS/CFT correspondence admit AdS vacuum solutions, but not all gauged
supergravities admit AdS vacua (see, for example, \cite{ss6}), and not
all supergravities which admit AdS vacua are gauged (see, for example,
\cite{pkt1}). We shall refer to those gauged supergravities that do
admit AdS vacua as ``AdS supergravities''.

  In this paper, we shall fill a gap in our knowledge of AdS
supergravities by constructing one in three dimensions with eight real
supersymmetries. We shall do this by means of a Scherk-Schwarz
reduction \cite{schsch} of the chiral $N=(1,0)$ supergravity in $D=6$.
The main motivation for our work is to make progress in finding the
still elusive supergravity theory that is expected to arise in the
AdS$_3\times S_3$ compactification of $(2,0)$ supergravity theory
\cite{dkss,db}, which in turn, emerges from type IIB string theory
reduced on $K3$ \cite{malda,ms}. As a first step in this direction,
here we construct the AdS$_3$ supergravity that is pertinent to the
AdS$_3\times S^3$ compactification of pure $N=(1,0)$ supergravity in
$D=6$ \cite{romans1}, which can also be embedded in heterotic string
compactifying on K3.  We do so by employing a Scherk-Schwarz reduction
on the $SU(2)$ group manifold. Before commenting further on our
results, let us review some facts about gauged and AdS supergravity
theories that can be obtained from consistent Kaluza-Klein reductions,
which will clarify the rationale behind our having chosen the
Scherk-Schwarz reduction scheme here.

   Many examples of gauged supergravities can be obtained by
consistent Kaluza-Klein sphere reduction of higher-dimensional
ungauged supergravities.\footnote{A {\it consistent} Kaluza-Klein
reduction is defined here as one where all the gauge bosons of the
isometry group $G$ of the compactifying manifold are retained in a
truncation keeping only a finite number of the lower-dimensional
fields, with the essential requirement that setting the truncated
fields to zero must be consistent with their own equations of motion.
Put another way, the reduction ansatz is consistent if all the
higher-dimensional equations of motion are satisfied as a consequence
of the equations of motion for the retained lower-dimensional
fields. It is only in very exceptional cases that such consistent
Kaluza-Klein reductions on compactifying spaces other than tori are
possible.} Notable cases include the $S^5$ reduction of type IIB
supergravity and the $S^7$ \cite{witnic} and $S^4$ \cite{vann1,vann2}
reductions of eleven-dimensional supergravity; these give rise to the
maximal gauged supergravities in $D=5$, $D=7$ and $D=4$
respectively.\footnote{To be precise, the consistency of the $S^5$
reduction has never been proven, and the reduction ansatz for the
$S^7$ example has not been fully explicitly exhibited.  The explicit
reduction ansatz for the $S^4$ case has been given,and its consistency
has been proven, modulo the assumption that the inclusion of quartic
fermion terms will not upset results established at lower order in
fermions.  In all cases, compelling circumstantial evidence implying
the consistency of the reductions has been found.} The reductions to
maximal supergravities are extremely complicated, especially in the
$S^5$ and $S^7$ cases.  Several examples giving gauged supergravities
with lesser supersymmetry have been worked out in complete detail (at
least in their bosonic sectors).  These include the half-maximal
supersymmetric gauged supergravities in $D=7$, 6, 5 and 4 dimensions
\cite{d7,d6,d5,d4}.

   The fact that the above Kaluza-Klein sphere reductions are
consistent is quite remarkable, and there is no general understanding
of why they work.  The consistency depends crucially on
``conspiracies'' between contributions from the metric and the other
fields in the higher-dimensional theories.  One might suspect from the
above examples that supersymmetry plays a crucial role, but in fact
this is somewhat misleading.  There are examples where supersymmetric
theories cannot be consistently reduced on spheres, and there are
examples where non-supersymmetric theories {\it can} be consistently
reduced on spheres.

    In fact a more universal characterisation of when a theory admits
a consistent sphere reduction can be given by first studying the
global symmetry of the theory when it is instead Kaluza-Klein reduced
on a torus of the same dimension as the sphere. This was discussed in
depth in \cite{s3s2red}.  The key point is that a generic theory
reduced on $T^n$ has a $GL(n,\R)$ global symmetry, and so its maximal
compact subgroup is $SO(n)$.  By contrast, the theory that one would
obtain by instead reducing on $S^n$ would have an $SO(n+1)$ gauge
group, and so by sending the gauge-coupling to zero there would have
to be at least an $SO(n+1)$ compact subgroup in the resulting global
symmetry group, contradicting the previous observation that
generically the maximal compact subgroup is only $SO(n)$.  Put another
way, if there were a consistent $S^n$ reduction then one would have to
be able to gauge an $SO(n+1)$ subgroup of the global symmetry group of
the $T^n$ reduction, and for a generic theory the toroidal reduction
does not yield a large enough global symmetry group.

   The only way in which a reduction on $S^n$ could be consistent is
therefore if there is actually an enhanced global symmetry group in
the reduction on $T^n$. This occurs only if there is some
``conspiracy'' between the contributions from the $T^n$ reduction of
metric and the other fields in the theory.  Such conspiracies are
indeed sometimes seen in supergravity reductions (including the
toroidal reductions of type IIB and eleven-dimensional supergravity),
and it is precisely these conspiracies that also allow the consistent
sphere reductions to work. However, as was shown in \cite{s3s2red},
there exist also examples of purely bosonic theories that also admit
consistent sphere reductions; a notable set of cases is provided by
the $S^3$ or $S^{D-3}$ reductions of the bosonic string effective
action in any dimension $D$.

   Since there exist AdS$_3\times S^3$ supersymmetric vacua of
the six-dimensional ungauged supergravities, it is natural to
enquire whether there might exist associated consistent $S^3$
reductions.  An extension of the arguments presented in
\cite{s3s2red} suggests that such consistent reductions are not
possible.\footnote{It is possible, however, to perform a different
consistent $S^3$ reduction that results in a gauged
three-dimensional supergravity which does not admit an AdS$_3$
vacuum solution, but instead a domain wall \cite{s3s2red}.  From the
six-dimensional point of view, this solution is the near-horizon limit of a
purely electric or purely magnetic string.} On the other hand, one
would expect that there should exist an AdS$_3$ gauged
supergravity, with a higher-dimensional origin, which would play
an important role in the AdS$_3$/CFT$_2$ correspondence.  This
would then be analogous to the examples in $4\le D\le 8$.

   Indeed there exists an alternative way of performing a consistent
Kaluza-Klein reduction on $S^3$, exploiting the fact that it is
isomorphic to the group manifold $SU(2)$.  This reduction, known
as Scherk-Schwarz reduction \cite{schsch}, has the merit that it
is guaranteed to be consistent when applied to any theory at all.
It does, however, give rise only to the gauge fields of $SU(2)$,
rather than the $SO(4)\sim SU(2)\times SU(2)$ that would have
arisen had there existed a consistent sphere reduction of the kind
we were describing above.  In this paper, we shall implement the
Scherk-Schwarz procedure for the case of an $S^3$ reduction of the
$N=(1,0)$ chiral supergravity in $D=6$.  This gives rise to a
gauged supergravity in three dimensions with an $SU(2)$ gauge
group that admits an AdS$_3$ vacuum solution.

   The gauged supergravity obtained here bears similarities to maximal
AdS supergravities in $D=5$ \cite{5dmax,pvn5}, $D=6$ \cite{6dmax} and
$D=7$ \cite{7dmax}. After we present our results, we shall comment on
these similarities in the concluding section.  It is worth pointing
out here, however, that our AdS$_3$ supergravity differs from other
gauged supergravities that have been constructed so far in three
dimensions \cite{at,it,ht,dkss2,n1,n2}, in which the Yang-Mills fields
are non-propagating since they belong to the supergravity multiplet
and are described solely by a Chern-Simons term.  Our model
corresponds to a fusion of non-propagating Poincar\'e supergravity
\cite{marsch,dw} with propagating fields originating from the
reduction of the graviton and 2-form field of $N=(1,0)$ supergravity
in $D=6$. In particular, the bosonic field content consists of a
dilaton; five scalars which parameterise the coset $SL(3,R)/SO(3)$; the
gauge fields of $SU(2)\sim SO(3)$, all of which originate from the
six-dimensional metric; and three vector fields which originate from
the six-dimensional 2-form potential. Thus altogether there are twelve
bosonic degrees of freedom, and by supersymmetry, twelve fermionic
ones. This is the same count of degrees of freedom that one finds in a
toroidal compactification of pure $N=(1,0)$ supergravity in $D=6$. We
shall compare this with the massless Kaluza-Klein spectra of the
AdS$_3\times S^3$ compactified $(1,0)$ and $N=(2,0)$ supergravities in
the concluding section.

   In this paper we shall also elaborate on the structure of the
scalar potential that arises in our model, describe the $U(1)$
truncation of the model, and we shall show how a charged black hole
solution can be obtained by performing a dimensional reduction of
the rotating self-dual string in the six-dimensional $(1,0)$
supergravity.

   The organisation of the paper is as follows.  In section 2 we begin
with a review of the Scherk-Schwarz $S^3$ reduction of a
$D$-dimensional metric, deriving expressions for the Ricci tensor in
$(D-3)$ dimensions.  We then specialise to the case of six-dimensional
$N=(1,0)$ supergravity, deriving the expressions for the reduction of
the self-dual 3-form field, and hence for the complete reduction of
the bosonic sector of the theory.  We show how the potential for the
$GL(3,\R)/SO(3)$ scalars can be expressed in terms of a
superpotential, and we derive a consistent truncation in which the
$SU(2)$ Yang-Mills fields are reduced to $U(1)$.

   In section 3 we extend the construction of the three-dimensional
gauged supergravity by obtaining the supersymmetry transformation rules
for the fermionic fields.  In section 4 we
consider some specific solutions of the three-dimensional supergravity,
and their six-dimensional interpretation in terms of rotating self-dual
strings.  The paper closes with conclusions and speculations in section 5.

\section{The bosonic sector}

   The bosonic sector of the $(1,0)$ six-dimensional supergravity
theory comprises the metric tensor $\hat g_{MN}$ and a 2-form
potential $\hat B_\2$ whose field strength $\hat H_\3 =d \hat
B_\2$ is self-dual. The six-dimensional bosonic equations of
motion are
\be \hat R_{MN} = \hat H_{MPQ}\, \hat H_N{}^{PQ}\,,\qquad d{\hat
*\hat H_\3}=0\,,\qquad {\hat *\hat H_\3}=\hat H_\3\,.
\label{d6bosonic} \ee

   We shall reduce the six-dimensional theory to three dimensions by
compactifying on a 3-sphere, viewed as the group manifold $SU(2)$.
The Kaluza-Klein reduction scheme that we employ will be the group
manifold reduction of Scherk and Schwarz \cite{schsch}, in which a
truncation to the set of all fields invariant under the left
action $G_L$ of the total isometry group $G_L\times G_R$ acting on
the group manifold $G$.  This truncation is guaranteed to be a
consistent one, since on group-theoretic grounds non-linear
products of the  retained $G_L$-singlet fields cannot act as
sources for the discarded $G_L$-non-singlet fields.

\subsection{Reduction of the metric}

   It is convenient to introduce the left-invariant $SU(2)$ 1-forms
$\sigma^\a$, which satisfy the Maurer-Cartan algebra
\be d \sigma^\a = -\ft12 f^\a{}_{\beta\gamma}\, \sigma^\beta\wedge
\sigma^\gamma\,,\label{cartan} \ee
where $f^\a{}_{\beta\gamma}$ are the $SU(2)$ structure constants.
For now we shall consider the case where we reduce on $SU(2)$ from
$(n+3)$ to $n$ dimensions.  The Kaluza-Klein metric reduction
ansatz will then be given by\footnote{In this paper we are using 
``supergravity conventions,'' in which the bosonic Lagrangian is
written with normalisations of the form
 $\sqrt{-g}\, (\ft14 R - \ft12 (\del\phi)^2 -\ft14 
(F_\2^i)^2 + \cdots)$.  For the convenience of readers who prefer the
customary $\sqrt{-g}\, (R - \ft12 (\del\phi)^2 -\ft14
(F_\2^i)^2 + \cdots)$ convention, we include a ``hidden'' appendix  
which repeats sections 2 and 4 in this notation.  It can be accessed by
deleting the ``$\backslash$end$\{$document$\}$'' in the Latex file at the
end of the references.}
\be d\hat s^2 = e^{2\a\, \phi}\, ds^2 + \fft{4}{g^2}\, e^{2\beta\,
\phi}\, h_{\a\beta}\, \nu^\a\, \nu^\beta\,,\label{metans} \ee
where $\phi$ is the ``breathing-mode'' scalar, $h_{\a\beta}$
denotes the remaining $n$-dimensional scalar fields (with the
symmetric tensor $h_{\a\beta}$ being unimodular), and $\nu^\a$ is
given by
\be \nu^\a \equiv \sigma^\a - g\,A^\a\,. \ee
Here $A^\a$ denotes the $SU(2)$ Yang-Mills potentials
corresponding to the right-acting $SU(2)$ isometry of the
3-sphere.  The constants $\a$ and $\beta$ in (\ref{metans}) will
be determined later.

   It will prove convenient to work in a vielbein basis, which we take
to be
\be \hat e^a = e^{\a\, \phi}\, e^a\,,\qquad \hat e^i = 2 g^{-1}\,
e^{\beta\, \phi}\,
 L^i_\a\, \nu^\a\,.\label{vielbein}
\ee
Here $e^a$ is a vielbein basis for the $n$-dimensional metric
$ds^2$, and $L^i_\a$ is a ``square root'' of $h_{\a\beta}$, and so
\be h_{\a\beta} = L^i_\a\, L^i_\beta\,,\qquad \det(L^i_\a)=1\,.
\ee

     Defining the Yang-Mills field strengths $F^\a = dA^\a + \ft12g\,
f^\a{}_{\beta\gamma}\, A^\beta\wedge A^\gamma$, we have:
\bea D\, F^\a &\equiv& dF^\a + g\,f^\a{}_{\beta\gamma}\,
A^\beta\wedge
F^\gamma=0\,,\nn\\
D\, \nu^\a &\equiv& d\nu^\a + g\,f^\a{}_{\beta\gamma}\,
A^\beta\wedge \nu^\gamma = -g\, F^\a -\ft12\,
f^\a{}_{\beta\gamma}\, \nu^\beta\wedge \nu^\gamma\,. \eea
It is also useful to define the Yang-Mills covariant exterior
derivative acting on the scalars $L^i_\a$:
\be D\, L^i_\a \equiv d L^i_\a - g\,f^\beta{}_{\gamma \a}\,
A^\gamma\,
 L^i_\beta\,.
\ee

  Using these expressions, we find from (\ref{vielbein}) that
\bea d\hat e^a &=& -\a\, e^{-\a\phi}\, \del_b\phi\, \, \hat
e^a\wedge \hat
e^b -\omega^a{}_b\wedge \hat e^b\,,\nn\\
d\hat e^i &=& e^{-\a\phi}\, (L^{-1})^\a_j\, (D_a L^i_\a)\, \hat
e^a \wedge \hat e^j + \beta \, e^{-\a\phi}\, \del_a\phi\, \hat
e^a\wedge \hat e^i - e^{(\beta-2\a)\phi}\, F_{ab}^i\,
\hat e^a\wedge \hat e^b \nn\\
&&- \ft14 g\,e^{-\beta\phi}\, T^{i\ell}\, \ep_{jk\ell} \,\hat
e^j\wedge
 \hat e^k\,,\label{deexp}
\eea
where we have defined
\be F^i_{ab} \equiv L^i_\a\, F^\a\,,\qquad T^{ij} \equiv L^i_\a\,
L^j_\a\,.\label{Tdef} \ee
(Note that $T^{ij}$ is $SU(2)$ covariant, despite superficial
appearances, since $\delta^{\a\beta}$ is an invariant tensor in
$SO(3)\sim SU(2)$.) From (\ref{deexp}), we calculate the
torsion-free spin connection $\hat \omega^A{}_B$, defined by
$d\hat e^A = -\hat\omega^A{}_B\wedge \hat e^B$ and
$\hat\omega_{AB}=-\hat\omega_{BA}$, finding
\bea \hat \omega_{ab} &=& \omega_{ab} + \a\, e^{-\a \phi}\,
(\del_b\phi\, \eta_{ac}\, \hat e^c - \del_a\phi\, \eta_{bc}\,
\hat e^c) +  e^{(\beta-2\a)\, \phi}\, F^i_{ab}\, \hat e^i\,,\nn\\
\hat \omega_{ai} &=& -e^{-\a\phi}\, P_{a\, ij}\,  \hat e^j -
\beta\, e^{-\a\phi}\, \del_a\phi\, \hat e^i +  e^{(\beta-2\a)\,
\phi}\,
  F^i_{ab}\, \hat e^b\,,\label{spincon}\\
\hat\omega_{ij} &=& e^{-\a\phi}\, Q_{a\, ij}\, \hat e^a + \ft14
g\, e^{-\beta\phi}\, (T^{k\ell}\, \ep_{ij\ell} + T^{j\ell}\,
\ep_{ik\ell} - T^{i\ell}\, \ep_{jk\ell})\, \hat e^k\,.\nn \eea
Here we have defined
\be P_{a\, ij} \equiv \ft12 [ (L^{-1})^\a_i\, D_a L^j_\a +
                           (L^{-1})^\a_j\, D_a L^i_\a]\,,
\qquad Q_{a\, ij} \equiv \ft12 [ (L^{-1})^\a_i\, D_a L^j_\a -
                           (L^{-1})^\a_j\, D_a L^i_\a]\,.
\ee

   The next step is to calculate the Ricci tensor $\hat R_{AB} = \hat
R^C{}_{ACB}$, which can in principle be done by first calculating
the curvature 2-forms $\hat\Theta^A{}_B = d\hat\omega^A{}_B +
\hat\omega^A{}_C\wedge \hat \omega^C{}_B = \ft12 \hat
R^A{}_{BCD}\, \hat e^C\wedge \hat e^D$.  This is quite an involved
calculation.  In practice, a simpler way to find the Ricci tensor
is to use an observation that was made in \cite{schsch}, which is
that the dimensional reduction of the Einstein Hilbert Lagrangian
$L=\hat {\rm e} \,\hat R$ is given, up to a total derivative,
by\footnote{It is crucial, in order to apply this argument, that
the reduction be a {\it consistent} one, meaning that the
equations of motion derived from the dimensionally-reduced action
coincide with those that follow from the dimensional reduction of
the higher-dimensional equations of motion.}
\be L = \hat {\rm e}\, ( \hat \omega_{A\, BC}\,\hat\omega^{C\, AB}
  + \hat\omega^A\, \hat\omega_A)\,,\label{lagexp}
\ee
where
\be \hat\omega_{AB}\equiv \omega_{C\, AB}\, \hat e^C\,,\qquad
\omega_A\equiv \eta^{BC}\, \hat\omega_{B\, CA}\,. \ee

   It is convenient at this point to make the following choices for the
constants $\a$ and $\beta$:
\be \a^2 = \fft{6}{(n-2)\, (n+1)}\,,\qquad \beta = -\ft13 \a\,
(n-2)\,. \label{albe} \ee
The second condition ensures that the lower-dimensional metric is
also in the Einstein frame, and the first condition simply sets
the scale for $\phi$ so that it has a canonically-normalised
kinetic term. After making these choices, we find after a simple
calculation from (\ref{spincon}) and (\ref{lagexp}) that the
higher-dimensional Einstein-Hilbert Lagrangian $L=\hat R\,
\sqrt{-\hat g}$ reduces to give
\bea \sqrt{-\hat g}\, \hat R &=& \sqrt{-g}\, \Big[ R  -  2
(\del\phi)^2 - (P_{a\, ij})^2
- e^{-\fft23\a\, (n+1)\, \phi}\, (F^i_{ab})^2 \nn\\
&& - \ft14 g^2\, e^{\ft23\a\, (n+1)\, \phi}\, (T^{ij}\, T^{ij}
-\ft12 T^2)\Big] +\hbox{total derivative}\,, \label{redlag} \eea
where $T\equiv T^{ii}$.

   From (\ref{redlag}), we can easily obtain the lower-dimensional
equations of motion for the metric, the Yang-Mills potentials, and
the scalar fields. These equations in fact match up with the $\hat
R_{ab}$, $\hat R_{ai}$ and $\hat R_{ij}$ vielbein components of
the higher-dimensional Ricci tensor.  The only subtlety is that
there are overall scalings to be determined, involving certain
specific powers of $e^\phi$, and that the components $\hat R_{ab}$
are actually formed from a linear combination of the
lower-dimensional Einstein equation and $\eta_{ab}$ times a
multiple of the trace of the scalar field equations.  These
scalings and combinations are easily determined by considering
special cases.  By this means, we therefore arrive at the
expressions for the components of the higher-dimensional Ricci
tensor with much less labour than by the direct approach via the
curvature 2-forms.  We find
\bea \hat R_{ab} &=& e^{-2\a\phi}\, [ R_{ab} -2 \del_a\phi\,
\del_b\phi - P_{a\, ij}\, P_{b\, ij} - \a\, \square\phi\,
\eta_{ab} - 2 e^{-\fft23\a\, (n+1)\, \phi}\, F^i_{ac}\, F^i_{bd}\,
\eta^{cd}]\,,\nn\\
\hat R_{ai} &=& -e^{\ft13\a\, (n-5)\, \phi}\, [{\cal D}^b\,
(e^{-\fft23\a\, (n+1)\, \phi}\, F^i_{ab}) + e^{-\fft23\a\, (n+1)\,
\phi}\, F^j_{ab}\, P^b{}_{ij} - \ft12 g\,\ep_{ijk}\,
  T^{k\ell}\, P_{a\, j\ell}]\,,\nn\\
\hat R_{ij} &=& -\ft12 e^{-2\a\phi}\, [ {\cal D}_a\, P^a{}_{ij} -
\ft23\a\, (n-2)\, \square \phi\, \delta_{ij} - 2 e^{-\fft23\a\,
(n+1)\, \phi}\, F^i_{ab}\, F^j_{cd}\, \eta^{ac}\,
\eta^{bd} \label{ricci}\\
&& - g^2\,e^{\ft23\a\, (n+1)\, \phi}\,(T^{ik}\, T^{jk} -\ft12 T\,
T^{ij}) + \ft12 g^2\,e^{\ft23\a\, (n+1)\, \phi}\,(T^{k\ell}\,
T^{k\ell}
 -\ft12 T^2)\, \delta_{ij}]\,,\nn
\eea
where we have now defined a derivative ${\cal D}_a$ that is
covariant not only with respect to general coordinate and
Yang-Mills transformations, but also it involves the composite
connection $Q_{a\, ij}$:
\be {\cal D}_a\, L^i_\a \equiv D_a\,  L^i_\a  + Q_{a\, ij}\,
L^j_\beta \,, \ee
(and similarly when acting on $P_{a\, ij}$ and $F^i$).  Note that
using ${\cal D}_a$, we have
\be (L^{-1})^\a_i\, {\cal D}_a \, L^j_\a = P_{a\, ij}\,. \ee

\subsection{Reduction of the 3-form}

   The self-dual 3-form in six dimensions can be written as $\hat H_\3
= \hat G_\3 + {\hat *\hat G_\3}$.  We find that the appropriate
reduction ansatz is given by taking
\be \hat G_\3 = \fft{8m}{g^3}\, \Omega_\3 + \fft2{g^2}
\ep_{\a\beta\gamma}\, B^\a\wedge \nu^\beta\wedge
\nu^\gamma\,,\label{gdef} \ee
where $m$ is a constant and $\Omega\equiv \nu^1\wedge \nu^2\wedge
\nu^3$. Dualising (\ref{gdef}) in the metric (\ref{metans}), we
find that
\be {\hat *\hat G_\3} = m\, e^{4\a\, \phi}\, \ep_\3 - \fft2{g}\,
e^{\ft43 \a\, \phi} h_{\a\beta}\, {* B^\a}\wedge \nu^\beta\,, \ee
where $h_{\a\beta} \equiv L^i_\a\, L^i_\beta$.  (Here we have used
equation (\ref{albe}), which for $n=3$ implies $\beta=-\ft13\a$
and $\a^2= \ft38$.) Noting that $d\Omega_\3 = -\ft12
g\,\ep_{\a\beta\gamma}\, F^\a\wedge \nu^\beta\wedge \nu^\gamma$,
we find that the Bianchi identity $d\hat H_\3=0$ implies the
equation
\be \left(D B^\a\right)L_\a^i - 2 m\, F^i + \ft12 g\,e^{\ft43\a\,
\phi}\, T^{ij}\, * B_j=0\,.\label{beqn} \ee

  We find that the vielbein components of the self-dual field strength
$\hat H_\3$ are given by
\bea &&\hat H_{abc} = m\, e^{\a\phi}\, \ep_{abc} \,,\qquad
\hat H_{abi} = -  e^{-\fft13 \a\phi}\, \ep_{ab}{}^c \, B_c^i\,,\nn\\
&& \hat H_{ijk} = m \, e^{\a\phi}\, \ep_{ijk}\,,\qquad \hat
H_{aij} = e^{-\fft13\a\phi}\, \ep_{ijk}\, B_a^k\,,\label{hcomp}
\eea
where we have defined $B^i\equiv L^i_\a\, B^\a$.

\subsection{Three-dimensional bosonic equations of motion and
Lagrangian}

   From (\ref{hcomp}), together with our expressions (\ref{ricci}) for
the Ricci tensor, we find that the six-dimensional bosonic
equations (\ref{d6bosonic}) imply the following three-dimensional
bosonic equations of motion.  First, from the $\hat R_{ij}$
components of the Einstein equation, we obtain the scalar
equations
\bea 
\a\, \square\phi &=& 6 m^2\, e^{4\a\phi} + \ft14 g^2\,
  e^{\ft83\a\phi}\, (T^{ij}\, T^{ij} -\ft12 T^2) + 2
e^{\ft43\a\phi}\, (B_a^i)^2 - e^{-\ft83\a\phi}\,
(F^i_{ab})^2\,,\nn\\
{\cal D}_a P^a{}_{ij} &=& g^2\,e^{\ft83\a\phi}\, [T^{ik}\, T_{jk}
-\ft12 T\, T^{ij} - \ft13 (T^{k\ell}\,
T^{k\ell} - \ft12 T^2)\, \delta_{ij}] \label{3dimeom}\\
&& +  8 e^{\ft43\a\phi}\, [B_a^i\, B_b^j\, \eta^{ab} - \ft13
(B_a^k)^2\, \delta_{ij} ] + 2 e^{-\ft83\a\phi}\, [F^i_{ab}\,
F^j_{cd}\, \eta^{ac}\, \eta^{bd} - \ft13 (F^k_{ab})^2 \,
\delta_{ij}]\,.\nn 
\eea

   From the $\hat R_{ai}$ components of the Einstein equation, we
obtain the Yang-Mills equation
\be {\cal D}^b(e^{-\fft83\a\phi}\, F^i_{ab}) =
-e^{-\fft83\a\phi}\, F^j_{ab}\, P^b{}_{ij} + \ft14 g^2\,
\ep_{ijk}\, T^{k\ell}\, P_{a\, j\ell} - 4 m\, e^{\ft43\a\phi}\,
B_a^i + 2 \ep_{ijk}\, \ep_{a}{}^{bc}\, B_b^j\, B_c^k\,. \ee
From the $\hat R_{ab}$ components of the Einstein equation we
obtain, after using the $\phi$ equation above to replace a
$\square\phi$ term,
\bea R_{ab} &=& 2 \del_a\phi\, \del_b\phi + P_{a\, ij}\, P_{b\,
ij} + 4  e^{\ft43\a\phi}\, B^i_a\, B^i_b + 2 e^{-\fft83\a\phi}\,
(F^i_{ac}\, F^i_{bd}\, \eta^{cd} - \ft12 (F^i_{cd})^2\,
\eta_{ab})\nn\\
&& + 4 m^2 \, e^{4\a\phi}\, \eta_{ab} + \ft14 g^2\,
e^{\ft83\a\phi}\, (T^{ij}\, T^{ij} -\ft12 T^2)\, \eta_{ab}\,.
\eea
Finally, there is the equation (\ref{beqn}) that came from $d\hat
H_\3=0$.

   We find that these equations of motions can be derived from the
3-dimensional Lagrangian
\bea {\cal L} &=& \ft14 R\, {*\oneone} - \ft12 {*d\phi}\wedge
d\phi - \ft14 {*P_{ij}}\wedge P_{ij} - e^{\ft43\a\phi}\,
{*B^i}\wedge B^i
- \ft12 e^{-\fft83\a\phi}\, {*F^i}\wedge F^i\nn\\
&& - m^2\, e^{4\a\phi}\, {*\oneone} - \ft1{16}
g^2\,e^{\ft83\a\phi}\, (T^{ij}\, T^{ij} -\ft12 T^2)\, {*\oneone}
+ {\cal L}_{CS}\,, \label{d3lag} \eea
where the Chern-Simons contribution ${\cal L}_{CS}$ is given by
\be {\cal L}_{CS} = -\fft2{g}\,DB^\a\wedge B^\a + \fft{8m}{g}\,
F^\a\wedge B^\a -
   \fft{8 m^2}{g}\, \omega_\3\,,
\ee
where 
\be
\omega_\3\equiv  
A^\a\wedge dA^\a +\ft13  \ep_{\a\beta\gamma}\,
A^\a\wedge A^\beta\wedge A^\gamma 
\ee
is the usual Chern-Simons 3-form for the $SU(2)$ Yang-Mills fields,
satisfying $d\omega_\3 = F^\a\wedge F^\a$.
Note that the Lagrangian is invariant under $m\longrightarrow -m$,
together with $B^\a\longrightarrow -B^\a$.  On the other hand, it is
invariant under $g\longrightarrow -g$ and $A^\a\longrightarrow -A^\a$
only if one also performs a parity or time-reversal transformation.

    Although one cannot directly take the $g\longrightarrow0$ limit in
the Lagrangian, it can clearly be done at the level of the
equations of motion.  This is analogous to the case of
seven-dimensional gauged supergravity, where the limit was
discussed in detail in \cite{s4red}.  The $g\longrightarrow 0$ limit
corresponds to a flattening of the reduction 3-sphere, which in
the truncation to the massless sector can be replaced by a torus.
The theory also admits a different limit, in which one instead
sends $m$ to zero.  As can be seen from (\ref{gdef}), this
corresponds to setting the 3-form flux on $S^3$ to zero.  The
maximally-symmetric vacuum solution would then be six-dimensional
Minkowski spacetime $d\hat s_6^2 = dx^\mu\, dx_\mu + dr^2 + r^2\,
d\Omega_3^2$, instead of AdS$_3\times S^3$.

   It is interesting to note that in the Scherk-Schwarz reduction of
ten-dimensional $N=1$ supergravity on $S^3$, one can consistently
truncate out the scalar fields that are parameterised by $L^i_\a$,
whilst retaining the $SU(2)$ Yang-Mills fields.  (This was proved
in \cite{dnphet1} for Scherk-Schwarz reductions of the low-energy
effective action of the bosonic string in any dimension, reduced
on any group manifold.  The truncation of the scalars is possible
provided that the vectors coming from the reduction of the 2-form
potential are equated to the Yang-Mills potentials.) By contrast,
in our present case we cannot consistently truncate the scalars
$L^i_\a$ without also truncating the $SU(2)$ Yang-Mills fields, as
can be seen from the scalar equations for ${\cal D}_a\, P^a_{ij}$
in (\ref{3dimeom}).  The key difference is that in the present
case the 3-form field strength in six dimensions is subject to a
self-duality condition.

\subsection{Superpotential for the three-dimensional theory}
\label{suppotsec}

    The scalar potential in (\ref{d3lag})
can be expressed in terms of a superpotential $W$.  To do so, it
is useful first to absorb the dilaton $\phi$ into $L^i_\a$, by
defining
\be \wtd L^i_\a \equiv e^{\ft23\a\, \phi}\, L^i_\a\,. \ee
The scalar sector of the Lagrangian (\ref{d3lag}) can then be
written as
\be {\cal L}_{\rm scal} =  - \ft14 {*\wtd P_{ij}}\wedge \wtd
P_{ij} -V\, {*\oneone}\,, \label{d3scalar} \ee
where the scalar potential is given by
\be V\equiv   m^2\, \det{\wtd T^{ij}} + \ft1{16} g^2\, (\wtd
T^{ij}\, \wtd T^{ij} -\ft12 \wtd T^2)\,, \ee
and $\wtd T^{ij} \equiv \wtd L^i_\a\, \wtd L^j_\a$.

   Introducing coordinates $\phi^I$ on the $GL(3,\R)/SO(3)$ scalar coset
manifold, a vielbein on the coset can be written as
\be V_I^{ij} = (\wtd L^{-1})^\a_i \, \bar D_I \wtd L^j_\a\,, \ee
where
\be
\bar D_I \wtd L^i_\a \equiv \fft{\del  \wtd L^i_\a}{\del\phi^I} + \bar
Q_I^{ij}\, \wtd L^j_\a\,,\qquad \bar Q_I^{ij} \equiv \ft12 \Big[
(\wtd L^{-1})^\a_i\, \fft{\del \wtd L^j_\a} {\del \phi^I}
    - (\wtd L^{-1})^\a_j\, \fft{\del \wtd L^i_\a}{\del \phi^I}\Big]\,.
\ee
In terms of the coset metric $G_{IJ} \equiv V_I^{ij}\, V_J^{ij}$,
the scalar Lagrangian (\ref{d3scalar}) can be written as ${\cal
L}_{\rm scal} = -\ft14 G_{IJ}\, {*d\phi^I}\wedge d\phi^J - V\,
{*\oneone}$.  The potential $V$ can be expressed in terms of a
superpotential $W$ as
\be V = G^{IJ}\, \fft{\del W}{\del \phi^I}\, \fft{\del W}{\del
\phi^J} -W^2\,, \ee
where we find that $W$ is given by
\bea W &=& -\sqrt2 m\, \sqrt{\det \wtd T^{ij}}
           + \ft{1}{4\sqrt2} \,g\,  \wtd T\,,\nn\\
&=&  - \sqrt2 m\, e^{2\a\, \phi} +
 \ft{1}{4\sqrt2} \,g\,  e^{\ft43\a\, \phi}\,
   T\,.\label{su2suppot}
\eea

    If $m$ has the same sign as $g$, (\ref{su2suppot}) has an extremum at
$T^{ij}=\delta^{ij}$, corresponding to the pure AdS$_3$
supersymmetric vacuum solution.  More generally, there are
solutions that break half the supersymmetry, which can be lifted
back to six dimensions where they correspond to the standard
self-dual string.  If, on the other hand, $m$ has the opposite
sign to $g$, then AdS$_3$ is not included among the solutions of
the associated first-order equations coming from $W$. There will
instead be a half-supersymmetric domain-wall solution, which can
be lifted back into six dimensions where it acquires an
interpretation as a disjoint ``interior'' branch of a
negative-mass self-dual string.  (This issue was discussed
extensively in \cite{clpdw}.)

\subsection{A consistent $U(1)$ truncation}\label{u1sec}

    It is straightforward to see that we can perform a consistent
truncation of the three-dimensional theory obtained in the
previous section, in which we set to zero two of the three
Yang-Mills potentials $A^\a$ and two of the $B^\a$ 1-forms (for
$\a=1$ and 2), and at the same time we truncate the 5 scalars in
the unimodular matrix $L^i_\a$ to a single diagonal scalar:
\be
L^i_\a = \pmatrix{e^{\gamma\, \varphi} & 0 & 0\cr
                     0 & e^{\gamma\, \varphi} & 0\cr
                     0 & 0 & e^{-2\gamma\, \varphi} }\,,\label{phispec}
\ee
where $\gamma\equiv 1/\sqrt{3}$. After establishing that the
truncation is consistent (by looking at the previous field
equations), we can then simply impose the truncation in the
Lagrangian (\ref{d3lag}).  This gives
\bea {\cal L} &=& \ft14 R\, {*\oneone} - \ft12 {*d\phi}\wedge
d\phi - \ft12 {*d\varphi}\wedge d\varphi -
e^{\ft43\a\phi-4\gamma\, \varphi}\, {*B}\wedge B
- \ft12 \, e^{-\fft83\a\phi-4\gamma\, \varphi}\, {*F}\wedge F\nn\\
&& - m^2\, e^{4\a\phi}\, {*\oneone} - \ft1{32} g^2\,
e^{\ft83\a\phi}\, (e^{-8\gamma\, \varphi} - 4 e^{-2\gamma\,
\varphi})\, {*\oneone} + {\cal L}_{CS}\,, \label{d3lag2} \eea
where the Chern-Simons contribution ${\cal L}_{CS}$ is given by
\be {\cal L}_{CS} = -\fft2{g}\,dB\wedge B + \fft{8m}{g}\, F\wedge
B -
   \fft{8m^2}{g}\, dA\wedge A\,.
\ee
Note that the Lagrangian is invariant under $m\longrightarrow -m$,
together with $B^\a\longrightarrow -B^\a$.  On the other hand, it
is invariant under $g\longrightarrow -g$ and $A^\a\longrightarrow
-A^\a$ only if one also performs a parity or time-reversal
transformation.

   The three-dimensional equations of motion following from this
truncated Lagrangian are:
\bea
&&\a\square\phi = 6 m^2\, e^{4\a\phi} + \ft18 g^2\,
e^{\ft83\a\phi}\, ( e^{-8\gamma\, \varphi} - 4 e^{-2\gamma\,
\varphi}) -e^{-\fft83\a\phi-4\gamma\, \varphi}\,
F^2 + \ft12 g^2\,e^{\fft43\a\phi-4\gamma\, \varphi}\, B^2\,,\nn\\
&&\gamma\, \square\varphi = -\ft1{12} g^2\,e^{\fft83\a\phi}\,
(e^{-8\gamma\, \varphi}-e^{-2\gamma\, \varphi}) - \ft1{3}
e^{-\ft83\a\phi-4\gamma\, \varphi}\,
F^2 -\ft13g^2\,  e^{\fft43\a\phi-4\gamma\, \varphi}\, B^2\,,\nn\\
&&d(e^{-\ft83\a\phi-4\gamma\, \varphi}\, {*F}) = -4 m\,
e^{\ft43\a\phi-4\gamma\, \varphi}\, {*B}\,,\nn\\
&&e^{\ft43\a\phi-4\gamma\,\varphi}\, {*B} = \fft{4m}{g}\, F
-\fft{2}{g}\,
dB\,,\nn\\
&&R_{ab} = \ft12 \del_a\phi\, \del_b\phi + 6 \del_a\varphi\,
\del_b\varphi +4 e^{\ft43\a\phi-4\gamma\, \varphi}\, B_a\, B_b +
2 e^{-\fft83\a\phi-4\gamma\, \varphi}\, (F^2_{ab} -\ft12
F^2\, \eta_{ab}) \nn\\
&&\phantom{XXX}+ 4 m^2\, e^{4\a\phi}\, \eta_{ab} + \ft18
g^2\,e^{\ft83\a\phi}\, ( e^{-8\gamma\, \varphi} - 4 e^{-2\gamma\,
\varphi})\, \eta_{ab}\,.\label{u1eom} \eea

    The corresponding truncation in the $SU(2)$ reduction ansatz is
given by
\bea d\hat s_6^2 &=& e^{2\a\phi}\, ds_3^2 + \fft{4}{g^2}\,
e^{-\fft23\a\phi}\, [e^{2\gamma\, \varphi}\, (\sigma_1^2 +
\sigma_2^2) +
e^{-4\gamma\, \varphi}\, (\sigma_3 -g\,A)^2]\,,\nn\\
\hat G_\3 &=& \fft{8m}{g^3}\, \sigma_1\wedge \sigma_2\wedge
(\sigma_3-g\,A) + \fft{4}{g^2}\,B\wedge \sigma_1\wedge \sigma_2\,.
\eea
The dual of $\hat G_\3$ is given by
\be {\hat *\hat G_\3} = m\, e^{4\a\phi}\, \ep_\3 - \fft2{g}\,
e^{\ft43\a\phi -4\gamma\, \varphi}\, {*B}\wedge
(\sigma_3-g\,A)\,. \ee

   It is also of interest to look for superpotentials from which the
scalar potential $V= m^2\, e^{4\a\phi} + \ft1{32} g^2\,
e^{\ft83\a\phi}\, (e^{-8\gamma\, \varphi} - 4 e^{-2\gamma\,
\varphi})$ can be derived. In this case, $V$ will be expressed as
\be V = \ft14 \Big(\fft{\del W}{\del \phi}\Big)^2 +
 \ft14 \Big(\fft{\del W}{\del \varphi}\Big)^2 - W^2\,.
\ee
We find that the following choices for $W$ are possible:
\bea W &=& - \sqrt2 m \, e^{2\a\, \phi} + \ft1{4\sqrt2} \,g\,
    e^{\ft43\a\, \phi}\, (e^{-4\gamma\, \varphi} + 2
e^{2\gamma\, \, \varphi})\,,\label{u1suppot1}\\
W &=& - \sqrt2 m \, e^{2\a\, \phi} + \ft{1}{4\sqrt2} \,g\,
    e^{\ft43\a\, \phi}\, (\pm e^{-4\gamma\,  \varphi} + 4
e^{-\gamma\, \varphi})\,.\label{u1suppot2} \eea
The choice in (\ref{u1suppot1}) correspond to the superpotential
given in (\ref{su2suppot}), specialised to the $U(1)$ truncation
given in (\ref{phispec}).  The choices in (\ref{u1suppot2}) do not
arise from the specialisation of (\ref{su2suppot}).  A particular
solution, if the minus sign is chosen, is the same AdS$_3$ vacuum.
But in general, the solutions of the first-order equations
associated with (\ref{u1suppot2}) are disjoint from those
associated with (\ref{u1suppot1}).


\section{The fermionic supersymmetry transformations}


The $D=6$, $(1,0)$   supergravity multiplet consists of the
vielbein, 2-form potential with self-dual field strength and a
gravitino which is symplectic Majorana-Weyl spinor in doublet
representation of the R-symmetry group $Sp(1)$. As is well known,
a manifestly covariant action containing these fields alone cannot
be written down due to the the self-duality condition. However,
the coupling of this multiplet to a tensor multiplet consisting of
a two-form potential with anti-self dual field strength, a dilaton
and anti-chiral symplectic-Majorana spinor does admit a Lagrangian
formulation. Indeed, the complete Lagrangian, field equations and
supersymmetry transformation rules for the coupled system have
been constructed in \cite{ns1}. Starting from these field
equations and transformation laws, we can obtain the corresponding
ones for the pure supergravity theory by setting the dilaton and
the tensor-multiplet spinor to zero, and imposing the exact,
supersymmetric self-duality condition
\be H^-_{ABC} + \ft18 {\bar\psi}_D\Gamma^{[D}
\Gamma_{ABC}\Gamma^{E]} \psi_E=0 \ . \la{sd1} \ee
One can show that the supersymmetric variation vanishes modulo the
gravitino field equation. To implement this self-duality
condition, we need to substitute

\be H_{ABC}= H^+_{ABC}-\ft18 {\bar\psi}_D\Gamma^{[D}
\Gamma_{ABC}\Gamma^{E]} \psi_E\ , \ee
everywhere $H_{ABC}$ occurs in the equations of motion and the
transformation laws.  

  In the present paper, we shall
just derive the fermionic supersymmetry transformation rules for the
Scherk-Schwarz reduced theree-dimensional theory; these suffice for testing 
the supersymmetry of three-dimensional bosonic solutions. In a later
paper, we shall present the entire fermionic three-dimensional
Lagrangian and transformation rules.

   The supersymmetry transformations obtained from \cite{ns1} by the
truncation of the tensor multiplet are given up to leading order
in fermions by
\bea 
\delta { e}_M{}^A &=& {\bar\epsilon}\,
\Gamma^A{\psi}_M\ ,\la{s1}\\[5pt]
\delta B_{MN}&=& -{\bar \epsilon}
\Gamma_{[M}{\psi}_{N]}\ ,\la{s2}\\[5pt]
\delta {\psi}_M &=& \left( {\nabla}_M  + \ft14 { H}^+_M{}^{AB}
\Gamma_{AB}\right) \epsilon\,,\la{s3} 
\eea
where $\nabla_M = \del_M + \ft14 \omega_M{}^{AB}\, \Gamma_{AB}$.  
The chiral truncation leading to
the a transformation rules have also been obtained in \cite{romans1}.

   To perform the reduction of the fermionic transformation rules, we
begin by making an ansatz for the reduction of the gravitino field. In
doing so, we shall make use of the original treatment of this problem
in \cite{schsch}, and \cite{ss8} where it has been studied further in
the context of $SU(2)$ reduction of $D=11$ supergravity. One technical
aspect of the reduction is the diagonalisation of the lower
dimensional gravitino and spinor kinetic terms. It is convenient to
treat the diagonalisation problem after performing the $SU(2)$
reduction, at the level of determining the field re-definitions in the
lower dimensional Lagrangian that will yield diagonal fermionic
kinetic terms. Thus, we begin with the ansatz
\be
{\hat\psi}_a (x,y)= e^{\delta\phi}\psi_a(x)\ , \quad\quad
{\hat\psi}_i(x,y)= e^{\delta\phi}\chi_i(x)\ , \la{a2}
\ee
where $\delta$ is a constant. To ensure that 
the gravitino kinetic term is canonical, with no 
dilaton prefactor, we must set 
\be 
\delta=-\ft12 \left[(n-1)\alpha+3\beta\right] 
\ee
when reducing from $n+3$ to $n$ dimensions.  In our case, with $n=3$, 
we therefore have
\be
\beta=-\ft13\alpha\ ,\quad\quad
\delta=-\ft12\alpha\ , \la{3d}
\ee
We also specify our conventions as follows:
\bea \Gamma^a &=& \gamma^a\times 1 \times \sigma^1\
,\quad\quad\quad\quad\quad\
\Gamma^i=1\times \gamma^i\times\sigma^2\\
C_{(6)}&=&(i\sigma^2)\times(i\sigma^2)\times \sigma^1\
,\quad\quad\  \ \Gamma_7=1\times 1\times\sigma^3\
,\\
\gamma^{abc}&=&\epsilon^{abc}\ , \quad
\gamma^{ijk}=i\epsilon^{ijk}\ ,\quad\ \ \
\{\gamma_a,\gamma_b\}=2\eta_{ab}\ , \quad
 \{\gamma_i,\gamma_j\}=2\delta_{ij}\ .
 \eea
We use the metrics $\eta_{AB}=(-+++++)$ and $\eta_{ab}=(-++)$.
Furthermore, $\epsilon^{012}=1=\epsilon^{345}$.  In our
convention, $\epsilon_{\mu\nu\rho}$ and $\epsilon^{\mu\nu\rho}$
are constant, and thus, $e\epsilon_{\mu\nu\rho}$ and
$e^{-1}\epsilon^{\mu\nu\rho}$ are tensors. 

   Although we shall not derive the complete fermionic sector of
the three-dimensional theory in the present paper, it is nonetheless 
useful to examine the structure of the fermionic kinetic terms, in order
to see how the diagonalisation of three-dimensional fermion fields 
should be performed.  To do this we can write down 
the Lagrangian of the six-dimensional tensor multiplet coupled supergravity
Lagrangian in which the dilaton field and the tensor multiplet
spinor are set to zero, namely
\be 
e^{-1} {\cal L} = \ft14 R -\ft1{12} H_{ABC}H^{ABC} -\ft12
{\bar \psi}_A\Gamma^{ABC}\psi_{BC}
-\ft1{24}{\bar\psi}^D\Gamma_{[D}\Gamma^{ABC}\Gamma_{E]}\psi^EH_{ABC}
\,. \la{lag1} 
\ee
After variation, one then needs to impose the self-duality condition
(\ref{sd1}).  Of course for our present purposes, only the gravitino
kinetic term is relevant.  Substituting (\ref{a2}) into the Lagrangian,
we obtain the three-dimensional kinetic terms
\be 
e^{-1}{\cal L}_{\rm kin} = -\ft12
{\bar\psi}_\mu\gamma^{\mu\nu\rho}{\cal D}_\nu\psi_\rho
-i{\bar\chi}_i\gamma^i\gamma^{\mu\nu}{\cal D}_\mu\psi_\nu
+\ft12 {\bar\chi}_i\gamma^{ij}\gamma^\mu{\cal D}_\mu\chi_j\,.
\ee
where have defined
\be
\psi_\mu= e_\mu^a\psi_a\ ,  \quad\quad
\Gamma_\mu=e_\mu^a\Gamma_a\,.
\la{def12}
\ee
As expected, the gravitino and spinor kinetic terms are mixed.
We have verified that they can be simultaneously be diagonalised
for any dimension $n$. In particular, the field redefinitions
which do the job for $n=3$ are given by
\bea
 \psi_\mu&=& \psi'_\mu -\ft{i}2 \gamma_\mu\gamma^k\chi'_k\
,\nn\\[5pt]
\chi_i&=& -\ft12\gamma^k\gamma_i\chi'_k\ .\la{fr} \eea
It follows that the inverse transformation is
\bea
\psi'_\mu&=&\psi_\mu+i\gamma_\mu\gamma^k\chi_k\ ,\nn\\
\chi'i&=&\gamma_{ij}\chi^j\ . \la{def12b} \eea
The kinetic terms become diagonal in terms of the primed fields,
and in particular the spinor kinetic term becomes
$-\ft12{\bar\chi}'^i\gamma^\mu{\cal D}_\mu\chi'_i$.

   The
supersymmetry transformations of the vielbein and gravitino in
pure $(1,0)$ supergravity in six dimensions are
\bea
{\hat e}_A{}^M\delta {\hat e}_{MB}&=& \hat{\bar\epsilon}\,
\Gamma_B{\hat\psi}_A \,,\la{ss1}\\[5pt]
\delta {\hat \psi}_A &=& {\hat\nabla}_A\,{\hat \epsilon}+\ft14
{\hat H}_A^{+\ CD}\Gamma_{CD}{\hat
\epsilon} \,.\la{ss2} \eea
As for the potential ${\hat B}_{MN}$, it is more convenient to
work with its field strength since its 3D field $B_\mu^\alpha$ it
gives rise to is related directly to the field strength ${\hat
H}_{abi}$ and ${\hat H}_{aij}$ as in \eq{hcomp}. From \eq{s2} we have
\be
\delta {\hat H}^+_{ABC} =-\ft32{\hat\nabla}_{[A}\Big(
\hat{\bar\epsilon} \Gamma_B{\hat\psi}_{C]}\Big)
-\ft14\epsilon_{ABCDEF} {\hat\nabla}^D \Big(\hat{\bar\epsilon}
\Gamma^E{\hat\psi}^F\Big)\,.
\ee

  To perform the $SU(2)$ reduction of the above transformation laws,
we begin by making an ansatz for the supersymmetry parameter
${\hat\epsilon}(x,y)$. Obtaining $\delta \psi_a(x)=\nabla_a
\epsilon(x)+\cdots$ requires the ansatz
\be 
{\hat\epsilon}(x,y)=e^{\alpha\phi/2}\epsilon(x)\,, 
\ee
where we have used \eq{3d}. This allows us to carry out the reduction of 
the gravitino transformation
rule \eq{ss2}; to leading order in
fermions, we find that this gives

\bea
\delta \psi_\mu&=& {\cal D}_\mu\epsilon -\ft14
ie^{-4\alpha\phi/3}\left( F_{\mu\nu}^i-2e^{2\alpha\phi}\gamma_\mu
B_\nu^i\right)\gamma^\nu\gamma_i\epsilon
\nn\\
&& +\ft12\alpha\gamma_\mu\gamma^\nu\partial_\nu\phi\epsilon -\ft12
me^{2\alpha\phi}\gamma_\mu\epsilon
,\la{sf6}\\[8pt]
\delta\chi_i &=& -\ft{i}2 \gamma^\mu\gamma^j\left(P_{\mu
ij}-\ft13\alpha\delta_{ij}\partial_\mu\phi\right)\epsilon +\ft14
e^{-4\alpha\phi/3}\left( F_{\mu\nu
i}-ee^{2\alpha\phi}\gamma_i\gamma^k\epsilon_{\mu\nu\rho}B^\rho_k\right)
\gamma^{\mu\nu}\epsilon\nn\\
&& +\ft{i}2 g
e^{4\alpha\phi/3}\left(T_{ij}-\ft12\delta_{ij}T\right) \gamma^j
\epsilon +\ft{i}2 m e^{2\alpha\phi}\gamma_i\epsilon \ . \la{sf7}
\eea
Note that an expected $T$-dependent term in the gravitino
transformation rule will emerge after performing the redefinition
\eq{fr}.

   After performing the redefinitions given in (\ref{def12}), we find
that the supersymmetry transformation rules for the diagonalised 
fermionic fields can be expressed as
\crampest
\bea
\delta \psi_\mu' &=& {\cal D}_\mu\, \ep + \sqrt2\, W\, \gamma_\mu\, \ep 
-i\, e^{\ft23\a\, \phi}\, B_\nu^i\, \gamma_\mu\, \gamma^\nu\, \gamma_i\, \ep
+ \ft{i}{4}\, e^{-\ft43\a\, \phi}\, (F^i_{\mu\nu}\, \gamma^\nu + 
F^i_{\rho\sigma}\, \gamma_\mu{}^{\rho\sigma})\, \gamma_i\, \ep\nn\,,\\
\delta \chi_i' &=& 
\ft{i}{2} \, V_I^{ij}\, \gamma_j\, (\gamma^\mu\, 
\del_\mu \phi^I -\sqrt2  G^{IJ}\, \fft{\del W}{\del \phi^J})\, \ep 
- e^{\ft23\a\, \phi}\, \gamma^\mu\, B_\mu^k\, \gamma_i\, \gamma_k\, 
\ep + \ft14 e^{-\ft43\a\, \phi}\, F_{\mu\nu}^j\, \gamma_{ij}\, \ep\,,
\eea
\uncramp
where $W$, $G_{IJ}$ and $\phi^I$ are the superpotential, scalar
sigma-model metric and target-space coordinates introduced in section
\ref{suppotsec}.

\section{Charged AdS$_3$ black hole}

   In this section, we shall show how a charged black hole solution in
the three-dimensional theory of section \ref{u1sec} can be
obtained, by performing a dimensional reduction of a rotating
self-dual string in the six-dimensional $(1,0)$ supergravity.  A
rotating dyonic string solution of six-dimensional $(1,1)$
supergravity was constructed in \cite{cvetlars}.  This involved
two angular momentum parameters $\ell_1$ and $\ell_2$, associated
with two commuting $U(1)$ factors in the $SO(4)$ rotation group
acting on 3-spheres in the four-dimensional transverse space of
the string.  There were also two parameters $\delta_1$ and
$\delta_2$ that characterised the electric and magnetic charges of
the string.  The configuration can be viewed as a solution in the
$(1,0)$ supergravity if one sets $\delta_1=\delta_2=\delta$, since
then the electric and magnetic charges become equal and
consequently the 3-form field strength becomes self-dual and the
dilaton decouples.  If the two angular momentum parameters are
also set equal, $\ell_1=\ell_2=\ell$, the rotating string solution
then fits within our $SU(2)$ reduction ansatz, and in fact it fits
within the $U(1)$ truncation of section \ref{u1sec}.  The metric
of this self-dual rotating string solution can be read off from
\cite{cvetlars}:
\bea d\hat s_6^2 &=& -H^{-1}\, (1-\fft{2 k}{r^2+\ell^2}) \, dt^2 +
H^{-1}\, dx^2 - \fft{H\, h^4\, \ell^2\, (r^2+\ell^2)} {(1+ h^2\,
\ell^2)}\, (c^2\,
dt + s^2\, dx)^2 \nn\\
&&+ \fft{H\, r^2\, (r^2+\ell^2)}{(r^2+\ell^2)^2 -2k\, r^2}\, dr^2
+ \ft14 H\, (r^2+\ell^2)\, [ \sigma_1^2+\sigma_2^2 + (1+h^2\,
\ell^2)\, (\sigma_3-\wtd A)^2]\,,\label{spinsolmet} \eea
where
\be H= 1+\fft{2k\, s^2}{r^2+\ell^2}\,,\qquad h^2 =\fft{2k}{H^2\,
(r^2+\ell^2)^2}\,,\qquad \wtd A= \fft{2h^2\, \ell}{1+h^2\,
\ell^2}\, (c^2\, dt + s^2\, dx)\,, \ee
and we have defined $c\equiv \cosh\delta$, $s\equiv \sinh\delta$.

    We find that the self-dual 3-form is given by
\bea \hat H_\3 \!\!\!&=&\!\!\! -\fft{k\, s\, c}{\sqrt2}\Big\{
\sigma_1\wedge\sigma_2 \wedge (\sigma_3-\wtd A) + [\wtd A -
\fft{2\ell}{H\,
(r^2+\ell)^2}\,(dt+dx)]\wedge \sigma_1\wedge\sigma_2 \label{spinsolh}\\
\!\!\!&&\!\!\! - \fft{4 r\, \ell}{H^2\, (r^2+\ell^2)^2}\,
(dt+dx)\wedge dr\wedge (\sigma_3-\wtd A) + \fft{8r}{H^2\, (1+h^2\,
\ell^2)(r^2+\ell^2)^2}\, dt\wedge dx\wedge dr\Big\}\,.\nn \eea

    Comparing the metric (\ref{spinsolmet}) and the field strength
(\ref{spinsolh}) with the reduction ansatz in section \ref{u1sec},
we obtain a three-dimensional AdS charged black hole solution,
given by
\bea ds_3^2 &=& \fft{g^6}{64} H^3\, (r^2+\ell^2)^3(1+h^2 \,
\ell^2)\Big\{
 -H^{-1}\, (1-\fft{2 k}{r^2+\ell^2}) \, dt^2 + H^{-1}\,
dx^2 \nn\\
&&- \fft{H\, h^4\, \ell^2\, (r^2+\ell^2)} {(1+ h^2\, \ell^2)}\,
(c^2\, dt + s^2\, dx)^2 + \fft{H\, r^2\,
(r^2+\ell^2)}{(r^2+\ell^2)^2 -2k\,
r^2}\, dr^2\Big\} \,,\nn\\
A &=& \fft{2h^2\, \ell}{g\,(1+h^2\, \ell^2)}\, (c^2\,
dt + s^2\, dx)\,,\label{d3sol}\\
B &=& {\sqrt2}\,\coth\delta\,  [-g\,A + \fft{2\ell}{H\,
(r^2+\ell^2)}\,
(dt+dx)]\,,\nn\\
e^{-2\a\phi} &=&\fft{g^6}{64}\, H^3\, (r^2+\ell^2)^3(1+h^2\,
\ell^2)\,,\qquad e^{-6\varphi} = 1+h^2\, \ell^2 \,,\nn \eea
where $m=-\sqrt2\,g\, \coth\delta$ and $g=\sqrt{2/(k\, s^2)}$.
One can indeed directly verify that this configuration satisfies
the three-dimensional equations of motion (\ref{u1eom}).

   In the extremal limit, which is obtained by sending $k$ to zero and
$\delta$ to infinity, keeping the charge parameter $Q=
k\,\sinh2\delta$ fixed, the solution becomes
\bea ds_3^2 &=& \fft{W}{Q^3}\Big[\fft{r^2+\ell^2}{W}(-dt^2 + dx^2)
- \fft{Q^2\,\ell^2}{W^3}\, (dt+dx)^2 + \fft{W\, r^2\, dr^2}{(r^2 +
\ell^2)^2}
\Big]\nn\\
A &=& \fft{\sqrt{Q^3}\, \ell\, (dt+dx)}{(r^2+\ell^2 +
Q)^2}\,,\quad B= \fft{2 \ell\, (r^2+\ell^2 + \ft12 Q)\, (dt+dx)}{
 (r^2+\ell^2 + Q)^2}\,,\nn\\
e^{-\fft23 \a\phi} &=& Q^{-1}\,(r^2 + \ell^2 + Q)\,,\qquad
\varphi=0\,, \eea
where $W=r^2 + \ell^2 + Q$.

     The charged AdS$_3$ black hole (\ref{d3sol}) also admits a decoupling
limit, namely
\be r=\epsilon\, \td r\,,\quad \ell=\epsilon\, \td \ell\,,\quad
k=\epsilon^4\, \td k\,,\quad \sinh 2\delta = \fft{Q}{\td k\,
\epsilon^2} \,,\ee
with $\epsilon$ taken to be small.  In this limit, the solution
becomes
\bea ds_3^2&=&\epsilon^2\Big[ - N^2\, dt^2 + \fft{1}{N^2}\, dr^2 +
r^2\Big(\fft{dx}{\sqrt{Q}} -
\fft{\ell^2}{\sqrt{Q}\, r^2}dt\Big)^2\Big]\,,\nn\\
A&=&\fft{\epsilon\, \ell}{\sqrt{Q}}\,(dt + dx)\,,\qquad
B=0\,,\qquad \phi=0\,,\qquad \varphi=0\,,\nn\\
g&=& \fft{2}{\epsilon\, \sqrt Q}\,,\qquad m = -
\fft{2\sqrt2}{\epsilon\, \sqrt Q}\,. \eea where
\be N^2=\fft{r^2}{Q} + \fft{2(\ell^2-k)}{Q} + \fft{\ell^4}{r^2\,
Q}\,. \ee
After performing the rescaling of three-dimensional fields and
coupling constants
\be g_{\mu\nu} = \ep^2\, \td g_{\mu\nu}\,,\quad A_\mu = \ep\, \wtd
A_\mu\,, \quad B_\mu = \ep\, \wtd B_\mu\,,\quad g = \ep^{-1}\, \td
g\,,\quad m= \ep^{-1}\, \td m\,, \ee
which leaves the equations of motion invariant, we obtain in the
$\ep\longrightarrow 0$ limit the three-dimensional BTZ black hole
metric \cite{btz}, with the mass $M$ and angular momentum $J$ given
by
\be M=\fft{2(k-\ell^2)}{Q}\,,\qquad J=\fft{2\ell^2}{\sqrt{Q}} \ee

    It is worth remarking that in our solution, the mass $M$ can have
either sign according to the relative values of $k$ and $\ell$.
It becomes negative when the non-extremality parameter $k$ goes to
zero. The BTZ black hole, which has a horizon, requires that $M$
be positive, and $M\ge |J|/\sqrt Q$.  In our parameterisation it
is then necessary that $2\ell^2 \le k \le \infty$.  It is rather
surprising that the extremal BTZ black hole arises from the
decoupling limit of the non-extremal rotating black self-dual
string with the non-extremality parameter $k=2\ell^2$.

      If on the other hand, we let $\ell\rightarrow {\rm i}\, \ell$,
the mass and angular momentum becomes
\be M=\fft{2(k+\ell^2)}{Q}\,,\qquad J=\fft{-2\ell^2}{\sqrt{Q}}\,.
\ee
The mass then satisfies the bound $M\ge |J|/\sqrt Q$ for all $k\ge
0$, with the extremal limit corresponding to $k=0$.


\section{Conclusion and discussion}


    In this paper we have carried out the Scherk-Schwarz reduction of
pure $N=(1,0)$ chiral six-dimensional supergravity.  Although the
reduction ansatz is guaranteed to be consistent, there is a subtlety
that the reduction can only be performed at the level of the equations
of motion, because the six-dimensional chiral theory cannot be
described in terms of a Lagrangian.  The resulting three-dimensional
theory, however, can be derived from a Lagrangian, and we have 
constructed this explicitly in the bosonic sector.

   The three-dimensional theory that we have constructed comprises the
metric, the Yang-Mills fields of $SU(2)$, three ``massive'' vectors in
the adjoint of $SU(2)$, six scalar fields in the coset
$GL(3,\R)/SO(3)$, and their fermionic partners.  The theory admits and
AdS$_3$ vacuum solution.  Although there exist many AdS$_3$ gauged
supergravities, ours is the only known example that has a
higher-dimensional string origin.  One feature of the Scherk-Schwarz
reduction that the breathing mode is part of the lower-dimensional
massless supermultiplet.  This contrasts with the situation in the
exceptional cases where a consistent Kaluza-Klein supergravity
reduction on a sphere such as $S^4$, $S^5$ or $S^7$ can be performed;
in these cases the breathing mode would be massive, but must be
excluded in the consistent reduction.  In particular, the inclusion of
the breathing mode in our Scherk-Schwarz reduction implies that we can
reduce the full six-dimensional self-dual string solution, and not
merely its AdS$_3\times S^3$ near-horizon limit, to give rise to a
domain wall in $D=3$.  Indeed, we obtained such a solution, with
rotation as well, by reducing the six-dimensional self-dual rotating
string.

   It is interesting to compare the spectrum of our model with that
arising in the linearised analysis of the AdS$_3\times S^3$
compactification of the $N=(1,0)$ supergravity, obtained in
\cite{dkss}. The massless bosonic sector of that reduction comprises
a dilaton, nine scalars and the gauge
fields of $SO(4)$, all of which originate from the metric, together
with an additional six vectors that come from the six-dimensional
2-form potential. This suggests the existence of a gauged supergravity
theory with $SO(4)$ gauge symmetry and a scalar sector describing a
$GL(4,R)/SO(4)$ coset, and admitting an AdS$_3$ vacuum. Note however
that the total bosonic degrees of freedom in this case will be 16, in
contrast to the 12 of our model obtained by $SU(2)$ reduction, and
thus it differs also from the count of 12 obtained from toroidal
compactification.  The model with $SO(4)$ gauge fields is therefore
not expected to arise from a consistent Kaluza-Klein reduction of the
pure $N=(1,0)$ supergravity in $D=6$. Moreover, it must differ from
the $SO(4)$ gauged supergravity implied by the results of
\cite{s3s2red}, which admits a domain wall, but not AdS$_3$, as a
solution.

   Of course, an AdS$_3$ supergravity with propagating $SO(4)$ gauge
sector may exist in its own right, notwithstanding the fact that it is
not expected to arise from a consistent Kaluza-Klein ansatz. This
latter statement would mean that caution is necessary when utilizing
such a theory in an AdS$_3$/CFT$_2$ correspondence, since the the
massive Kaluza-Klein modes may have to be taken into account in this
case. In fact, the situation is similar to that encountered in the
$T^{1,1}$ compactification of the type IIB string, for which the
linearised analysis yields a minimal gauged supergravity coupled to
matter in $D=6$. A supergravity theory with the same massless spectrum
does exist in its own right, but it cannot arise from a consistent
truncation of the massive Kaluza-Klein modes in the $T^{1,1}$
reduction of type IIB supergravity, since there can be no such consistent
reduction ansatz \cite{hoxmarpop}.

   A model of great interest in the AdS$_3/CFT_2$ context is the
AdS$_3\times S^3$ compactification of the $N=(2,0)$ supergravity
in $D=6$.  In this case, the theory has 16 real supersymmetries, both
in $D=6$ and $D=3$, and it was determined in \cite{dkss} that the
propagating massless Kaluza-Klein spectrum consists of 21
hyper-multiplets and a special $32+32$ vector multiplet which consists
of $SO(4)$ Yang-Mills fields, 6 additional vector fields and 26
scalar fields, which include a dilaton and other scalars in various
representations of the $SO(4)_{\rm local}\times SO(4)_{\rm global}$ groups
involved. The presence of this multiplet has not been emphasized
in the literature so far, but it is clearly of great relevance to
the ultimate construction of the AdS supergravity theory that
describes the full massless spectrum. The model presented in
\cite{n1} describes the coupling of arbitrary number of
hyper-multiplets to an AdS$_3$ supergravity with $16$ real
supersymmetries, but it lacks the coupling of the propagating vector
multiplet mentioned above, and as such the problem of constructing
the desired supergravity theory still remains open. It is tempting
to conjecture that the $26$ scalars in the model parameterise an
$E_6/F_4$ coset. This problem is currently under investigation
\cite{bss}.

    There are other interesting aspects of gauged or AdS supergravities
in $D=3$, having to do, for example, with the understanding of the
relation between the models constructed in \cite{it} and
\cite{dkss2,n1,n2}. Ultimately, progress in this area ought to
lead to a better understanding of various interesting
AdS$_3$/CFT$_2$ issues as well. Of course, the  supersymmetric field
theories in $D=3$ may have  a number of other applications too,
including  some aspects of brane dynamics. It is interesting that
after years of progress in more complicated higher-dimensional
supergravities, it is relatively recently that the interest in $D=3$
supergravities has increased.  It is becoming increasingly
clear that this is a very rich area with a wealth of phenomena
still waiting to be uncovered.

\section*{Acknowledgement} We are grateful to Eric Bergshoeff and 
Per Sundell for useful discussions.

\newpage


\end{document}